\def\year{2021}\relax
\documentclass[letterpaper]{article} 
\usepackage{amsmath}
\usepackage{aaai21}  
\usepackage{times}  
\usepackage{helvet} 
\usepackage{courier}  
\usepackage[hyphens]{url}  
\usepackage{graphicx} 
\usepackage{natbib}  
\usepackage{caption} 
\usepackage{booktabs}

\title{Emergent Instabilities in Algorithmic Feedback Loops}

 \author{
     Keith Burghardt, Kristina Lerman  \\
 }
\affiliations{USC Information Sciences Institute\\
4676 Admiralty Way, Marina del Rey, CA, USA\\
\{keithab, lerman\}@isi.edu}
\begin{document}
\maketitle

\begin{abstract}
Algorithms that aid human tasks, such as recommendation systems, are ubiquitous. They appear in everything from social media to streaming videos to online shopping. However, the feedback loop between people and algorithms is poorly understood and can amplify cognitive and social biases (algorithmic confounding), leading to unexpected outcomes. In this work, we explore algorithmic confounding in collaborative filtering-based recommendation algorithms through teacher-student learning simulations. Namely, a student collaborative filtering-based model, trained on simulated choices, is used by the recommendation algorithm to recommend items to agents. Agents might choose some of these items, according to an underlying teacher model, with new choices then fed back into the student model as new training data (approximating online machine learning). These simulations demonstrate how algorithmic confounding produces erroneous recommendations which in turn lead to instability, i.e., wide variations in an item's popularity between each simulation realization. We use the simulations to demonstrate a novel approach to training collaborative filtering models that can create more stable and accurate recommendations. Our methodology is general enough that it can be extended to other socio-technical systems in order to better quantify and improve the stability of algorithms. These results highlight the need to account for emergent behaviors from interactions between people and algorithms.  
\end{abstract}



\begin{figure*}
  \includegraphics[width=\textwidth]{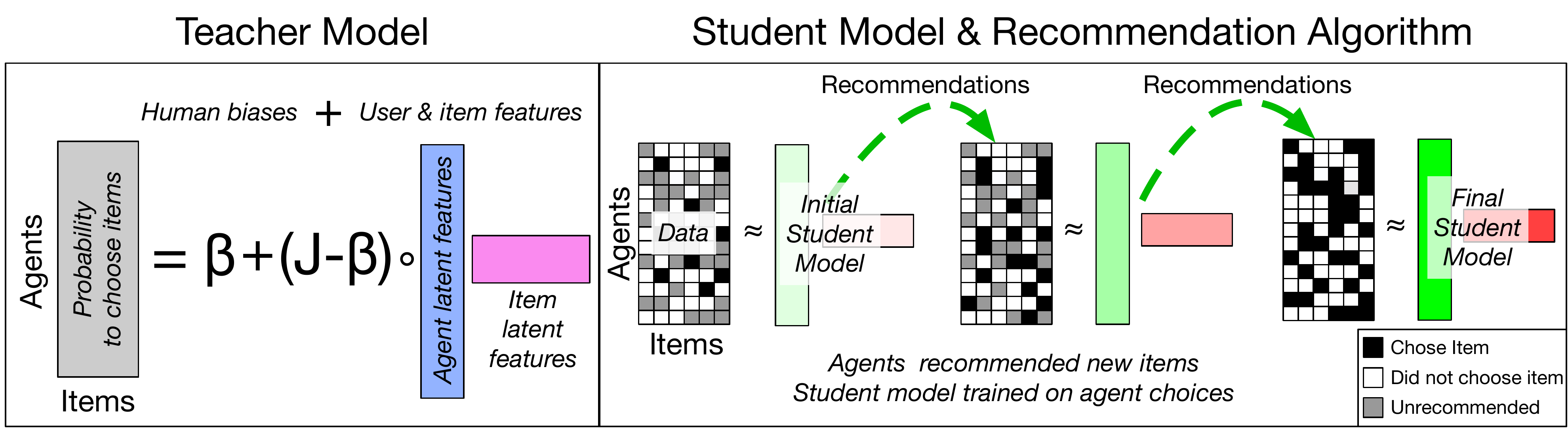}
  \caption{Framework for simulating recommendation systems. Left panel: a ground truth (teacher) model that simulates how agents choose items. This model is composed of a probability matrix that agents choose an item recommended to them for reasons besides the intrinsic item features \cite{Salganik2006,Burghardt2020} plus a matrix that is the product of two low-rank matrices to approximate how user and item latent features affect what items are chosen \cite{funk2006netflix,Bell2007,Koren2009,Portugal2018}. Right panel: the recommendation algorithm. A collaborative filtering (student) model approximates the user-item matrix as the product of two low-rank matrices \cite{funk2006netflix,Bell2007,Koren2009,Portugal2018}, known as matrix factorization, while the learning framework consists of recommending new items, recording what user-item pairs have been chosen, and retraining the student model on all collected data at the end of the timestep.}
\label{fig:teaser}
\end{figure*}

\section{Introduction}
We interact with computer algorithms throughout the day. Algorithms guide us to our destinations, finish our sentences in emails, automate business processes and decisions in healthcare, and recommend movies and music for us to enjoy.  On social media platforms, algorithms select  messages for our news feed and find new accounts for us to follow. Despite algorithms' ubiquity and broad impact, the interaction between algorithms and people within socio-technical systems is still poorly understood, especially when algorithms learn from data based on past predictions \cite{Sinha2016}. 

We have some evidence, however, that this interaction can have unexpectedly negative consequences, such as pushing people into filter bubbles \cite{Ge2020,Sirbu2019} or reducing the diversity of recommended content in online platforms \cite{Chaney2018,Mansoury2020}. Something observed in crowdsourcing systems, but under-explored in other systems, is how the interplay between people and algorithms can stochastically make some options very popular or unpopular (known as \textit{instability}), and the popularity of items has little relation to the items people prefer~\cite{Salganik2006,Burghardt2020}.  Mitigating this problem in crowdsourcing systems is an ongoing struggle but, until now, it was unknown if this instability extends to other socio-technical systems, such as personalized recommendation. This feedback loop could make many items or videos more popular than better ones and generate more unpredictable trends. These emergent problems are bad for recommendation systems because they will recommend items or content users are less likely to want, reducing revenue. The unpredictability is a new challenge for businesses or content creators, who would be less certain about what items or content will be the next big thing. These problems lead us to explore two research questions in this paper:
\begin{enumerate}
\item[\textbf{RQ1}] How can we measure the stability of non-crowdsourced socio-technical systems?
\item[\textbf{RQ2}]  Can we improve the stability and performance of a socio-technical system?
\end{enumerate}

We address these gaps in knowledge by systematically studying the complex dynamics of an algorithmically-driven socio-technical system, and focus on recommender systems. There are many systems we could explore, such as predictive policing \cite{Ensign2018}, or bank loans \cite{DAmour2020}, in which a feedback loop between people interacting with algorithms and the algorithms trained on these interactions could unfairly benefit some people over others.  Recommender systems are chosen because they are a key component of social media platforms, e-commerce systems \cite{Ge2020}, and are used by many music and video streaming services \cite{Bell2007,Schafer2007}. In order to better understand the feedback loop, known as algorithmic confounding, in recommendation systems, we create a teacher-student framework that approximates a continuously trained recommendation system \cite{Lampinen2018}, as shown in Fig.~\ref{fig:teaser}. The left panel of Fig.~\ref{fig:teaser} shows the ground truth \emph{teacher model}, which models human choices within recommendation systems using agents. These agents stochastically choose items according to intrinsic features of items and preferences of users (modeled as a factorized matrix, which fits empirical data well \cite{funk2006netflix,Bell2007,Koren2009,Portugal2018}), as well as human biases (such as choosing items at random, a heuristic seen in crowdsourcing \cite{Burghardt2020}). This model gives each user-item pair a unique probability to be chosen by a simulated agent when the item is personally recommended by the algorithm. We can alternatively interpret ``choosing'' items as agents rating something as good or bad (a goal of some recommendation systems \cite{funk2006netflix}), but we will stick to choosing items for consistency. 

We train a separate \textit{student model} from agent's binary decisions (right panel of Fig.~\ref{fig:teaser}) and use it to make new recommendations to agents. The student model is a separate factorized matrix that estimates what items each agent is likely to choose. The system records whether agents do or do not choose recommended items and the data is used to further fine-tune the student model. We have agents interact with items exactly once in order to model agents preferring new content over old \cite{Sinha2016} although this assumption can be easily modified in the code we provide, and should not affect our results. 

Our results show that recommending items that agents are predicted to like most leads to item popularity instability, where the same item can be very popular or unpopular in different simulation realizations. We use the simulation to test an alternative recommendation strategy in which random items are sometimes recommended. This new strategy improves stability and model accuracy as well as mean item popularity (a proxy for purchases or view counts of videos) at a given time. Moreover, a side-benefit of this strategy is that it forces the algorithm to recommend diverse content, which could reduce recommendation filter bubbles.

To summarize, our contributions are the following:
\begin{enumerate}
    \item We develop a novel framework to evaluate the stability of model training \footnote{The simulation code can be found here: https://github.com/KeithBurghardt/RecSim}.
    \item We quantify the stability of different recommendation algorithms to algorithmic confounding.
    \item We provide a simple recommendation algorithm strategy that improves accuracy and stability.
\end{enumerate}
These results demonstrate that personalized recommendation systems can produce exceedingly unstable recommendations. While the simulation is an idealized system, it gives new insight into why the systems work, why they sometimes fail, and algorithm strategies to mitigate their shortcomings.

\begin{table*}[tbh!]
    \centering
    \begin{tabular}{|p{3.5cm}|p{13cm}|}
    \hline
    {\bf Terminology} & {\bf Definition} \\\hline\hline
\textit{Personalized recommendations} & Recommendations unique to each person, in contrast to crowdsourced ranking, where the same recommendations are seen by everyone\\\hline
\textit{Algorithmic confounding} & The positive feedback loop between model and people in recommendation systems\\\hline
\textit{Simulation} & The evolution of the recommendation system from the initial condition until agents choose all items\\\hline
\textit{Agent} & The simulated user in our model\\\hline
\textit{Items} & What the recommendation algorithm recommends and agents choose\\\hline
\textit{User-item matrix} & The Boolean matrix of items each agent has been recommended, that are (1) or are not (0) chosen (these could be alternatively interpreted as binary ratings) \\\hline
\textit{Matrix factorization} & Algorithms that approximate a matrix as the product of two lower-rank matrices\\\hline
\textit{Recommendation system} & The recommendation algorithm along with the agents who interact with it\\\hline
\textit{Recommendation algorithm} & The matrix factorization-based student model and algorithm to recommend items to agents\\\hline
\textit{Teacher model} & The model approximating how agents interact with items. Namely the likelihood any item is selected\\\hline
\textit{Student model} & The recommendation model that approximates the available user-item matrix with matrix factorization in order to approximate how agents choose unseen items\\\hline
\textit{Teacher-student framework} & The framework in which a student model tries to reconstruct a teacher model based on the teacher model’s output\\\hline
\textit{Instability} & Sensitivity of recommended items to initial conditions\\\hline
\textit{Popularity} & How many agents have chosen an item at a given timepoint\\\hline
\textit{Quality} & The overall probability an item is select. These are elements of the teacher matrix\\\hline
    \end{tabular}
    \caption{Terminology definitions.}
    \label{tab:terminology}
\end{table*}

\section{Related Work}

\paragraph{Recommendation systems} 
There has been significant research into improving recommendation systems. Many methods exist to recommend everything from game characters \cite{conley2013does} to educative content \cite{Tan2008} to movies \cite{biancalana2011context}, and are often based on collaborative filtering (recommending based on the behavior of similar people), content filtering (recommending similar content), or a combination of both \cite{Balabanovic1997,Schafer2007,Portugal2018}. Collaborative filtering, which the present paper simulates, can use a number of different models from more K-means and ensemble-based methods to neural networks \cite{He2017,Bell2007,kim2016convolutional}. 

A popular and accurate recommendation model is matrix factorization, in which the sparse matrix of users-items pairs, $\mathbf{R}^\text{data}$, is approximated as the product of two lower-rank matrices $\approx \mathbf{P'} \mathbf{Q'}^T$ \cite{Koren2009}. Throughout the paper, matrices are $\mathbf{bold}$ while elements within a matrix are italicized. The intuition behind matrix factorization is that users and items may individually have latent features that make users more likely to pick one item (such as watching an action movie) over another (such as watching a romantic comedy). There has been significant interest in matrix factorization both due to its performance \cite{Bell2007,kim2016convolutional}, and relative ease to analyze theoretically \cite{Lesieur2017}. This method is often used in conjunction with other models, but for simplicity we model matrix factorization alone in the present paper.

\paragraph{Algorithm biases}
Training on biased data, a common practice in recommendation systems, can enhance biases, leading to greater unfairness and more mistakes  \cite{Ensign2018,DAmour2020,Jabbari2017,Joseph2016,angwin2016machine}. This is known as algorithmic bias or algorithmic confounding in recommendation systems \cite{Mansoury2020,Chaney2018,Sinha2016}. This bias might create filter bubbles that enhance polarization \cite{Sirbu2019,Bessi2016}. 

\paragraph{Ranking instability in crowdsourcing} A large body of literature has explored the behavior of crowdsourcing systems. In contrast to recommendation systems that personalize content, in crowdsourcing systems all users see the same content. These systems aggregate decisions of many people to find the best items, typically by ranking them.  Examples include StackExchange, where users choose the best answers to questions, and Reddit, where users choose the most interesting stories for the front page.  Past work has shown that ranking, especially by popularity, creates feedback loops that amplify human biases affecting item choices, such as choosing popular items or those they see first, rather than high-quality items  
\cite{lerman14as,MyopiaCrowd}. Recent literature has also identified instabilities in crowdsourced ranking \cite{Burghardt2020,Salganik2006}, in which the crowdsourced rank of items are strongly influenced by position and social influence biases. As a result, the emergent popularity of mid-range quality content is both unpredictable and highly uneven, although the best (worst) items usually end up becoming most (least) popular~\cite{Salganik2006,Burghardt2018}. Along these lines, Burghardt et al. (\citeyear{Burghardt2020}) developed a model to explain how the better item in a two-item list was not guaranteed to become highest ranked, which implies good content is often harder to spot unless the ranking algorithm controls for these biases. Finally, content recommended by Reddit had a poor correlation with user preferences \cite{Glenski2018}, suggesting factors including algorithmic confounding have produced poor crowdsourced recommendations.


\paragraph{Reinforcement Learning} Reinforcement learning is the algorithmic technique of interacting with an environment with a set of actions and learning what actions maximize cumulative utility \cite{Kaelbling1996}. A number of reinforcement learning methods exist, from genetic algorithms and dynamic programming \cite{Kaelbling1996} to deep learning-based algorithms \cite{Arulkumaran2017}. The typical goal is to initially explore the space of actions and then exploit actions learned that can optimize cumulative utility. Personalized recommendations are a natural fit for reinforcement learning because users choose items sequentially, and the metrics to optimize, e.g., money spent or videos watched, can easily be interpreted as a cumulative utility to optimize. A growing body of literature has shown how reinforcement learning algorithms can learn a sequence of recommendations to increase the number of items bought or content viewed \cite{Taghipour2007,Lu2016,Arulkumaran2017,Afsar2021}. These algorithms are often trained on data of past users interacting with the system, which could lead to algorithmic confounding. The framework within the present paper can be extended in the future to measure algorithmic confounding within reinforcement learning-based recommendation systems and help companies develop new techniques to better train these models in online settings.

\paragraph{Our novelty} The present paper contrasts with previous work by developing a teacher-learner framework to model and better understand the interaction between recommendation algorithms and users. Furthermore, we use these findings to demonstrate how item instability can be a vexing feature of recommendation systems, in which popular or highly recommended items may not strongly correlate with items agents prefer. Finally, we provide a novel reinforcement learning-inspired approach to better train collaborative filtering models, which can improve stability and recommendation accuracy.

\begin{table}[tbh!]
    \centering
    \begin{tabular}{|p{1.1cm}|p{6cm}|}
    \hline
    {\bf Symbol} & {\bf Definition} \\\hline\hline
    $\mathbf{R}^{\text{data}}$ & User-item matrix\\\hline
    $\mathbf{R}^{\text{teacher}}$ & Teacher model\\\hline
    $\mathbf{R}^{\text{student}}$ & Student model\\\hline
    $\boldsymbol{\beta}$ & The teacher model probability to choose items independent of their features \\\hline
    $\beta$ & Scalar value of all elements in $\boldsymbol{\beta}$ (free parameter between 0 and 1)\\\hline
    $\boldsymbol{J}$ & All-ones matrix\\\hline
    $\circ$ & Hadamard product (element-by-element multiplication between matrices)\\\hline
    $\mathbf{P}$, $\mathbf{Q}$ & The latent features of users ($\mathbf{P}$) and items ($\mathbf{Q}$) in the teacher model \\\hline
    $\mathbf{P'}$, $\mathbf{Q'}$ & The estimated latent features of users ($\mathbf{P'}$) and items ($\mathbf{Q'}$) in the student model \\ \hline
    $k$ & Number of latent features in the teacher model ($k=4$)\\\hline
    $k'$ & Number of latent features in the student model ($k'=5$)\\\hline
    $n$ & Number of agents ($n=4000$) \\\hline
    $m$ & Number of items ($m=200$) \\\hline
    $A_{ij}$ & $ij^{th}$ element in a matrix $\mathbf{A}$\\\hline
    $T$ & Timestep (value from $1$ to $m$)\\\hline
    \end{tabular}
    \caption{Symbol definitions.}
    \label{tab:symbol}
\end{table}

\section{Methods}

We introduce the teacher-student modeling framework, simulation assumptions, and the recommendation algorithm strategies for student model training. Terminology referenced this section is available in Table~\ref{tab:terminology}, and symbols can be referenced in Table~\ref{tab:symbol}.

\subsection{Outline of our approach}
We analyze recommendation systems using simulations that capture the essence of recommendation algorithms while being simple enough to make generalizable conclusions. Model simplifications include:
\begin{enumerate}
    \item A recommendation algorithm recommends $r=1$ items to each agent before retraining on past data.
    \item We assume agents are roughly the same age within the system.
    \item Agents make binary choices, alternatively interpreted as ratings (upvote or downvote) \cite{Trnecka2021}.
    \item Agent decisions follow the teacher model seen in the left panel of Fig.~\ref{fig:teaser}.
    \item Agent choices are the same regardless of the order items are offered.
    \item Regardless of agent choice, the item is not recommended again to that agent. This captures how old items may have a lower utility to users than novel items \cite{Chaney2018}.
\end{enumerate}
Additional realism can be built into the simulation in the future; the present work can be viewed as a proof-of-concept.

\subsection{Teacher-student framework}

The recommendation system simulation has two components: a \textit{teacher model}, $\mathbf{R}^{\text{teacher}}$, which models agent decisions, and a \textit{student model}, $\mathbf{R}^{\text{student}}$, which models the recommendation engine. The student model is trained on a matrix of recommended items agents have or have not chosen, $\mathbf{R}^{\text{data}}$. The joint teacher-student framework has the following benefits: (1) the teacher model encodes agent preferences and can be made arbitrarily complex to improve realism, (2) the student model prediction can be directly compared to the teacher model ground truth, and (3) we can explore counterfactual conditions of how the recommendation system would behave if agents chose a different set of items.

The left panel of Fig.~\ref{fig:teaser} shows the teacher model, which assumes that agents choose items stochastically according to a probability matrix that models both human biases and intrinsic preferences. The teacher model is  
\begin{equation}
\mathbf{R}^{\text{teacher}} = \boldsymbol{\beta} + (\boldsymbol{J}-\boldsymbol{\beta}) \circ \mathbf{P}\mathbf{Q}^T    
\end{equation}
where $\boldsymbol{\beta}$ is a matrix representing the probability a user will pick a given item regardless of its intrinsic qualities, $\boldsymbol{J}$ is an all-ones matrix, $\circ$ is the Hadamard product, and $\mathbf{P}$ and $\mathbf{Q}$ are both low-rank matrices. This last term approximates how agents choose items due to their intrinsic preferences for items with particular latent features. The teacher model is similar to previous models of human decisions in crowdsourced systems \cite{Burghardt2020}, where agents stochastically choose items due to intrinsic qualities or at random due to human biases. The biases in real systems could vary in intensity for different scenarios, so we keep $\boldsymbol{\beta}$ as a set of free parameters. For simplicity, we initially set the matrix $\boldsymbol{\beta} = \beta \mathbf{J}$, where $\beta$ is a scalar. For robustness, we compare our results to the case when the $\boldsymbol{\beta}$ matrix are probabilities distributed uniformly at random between 0 and 1 (random $\boldsymbol{\beta}$ condition).

To further simplify the simulation, we let $\mathbf{P}$ and $\mathbf{Q}$ be rank-$k$ whose elements are uniformly distributed between 0 and $1/k$. This ensures that, after the rank-$k$ matrices are multiplied, the probabilities are always positive definite and less than 1 (on average 0.25, which avoids highly imbalanced datasets), while otherwise making minimal assumptions about the matrix element values. In this model, we arbitrarily choose $k=4$ to ensure that $k\ll n,~m$ as is typical in other recommendation systems \cite{Bell2007,funk2006netflix}.

The student model approximates the user-item matrix, $\mathbf{R}^{\text{data}}$, with matrix factorization
\begin{equation}
    \mathbf{R}^{\text{student}}=\mathbf{P'}\mathbf{Q'}^T
\end{equation} 
shown in the right panel of Fig.~\ref{fig:teaser}. Matrix factorization assumes that agent choices are best explained by a smaller number of latent factors (agent preferences) that can be learned from their observed choices, as we assume in the teacher model. The model's output is the expected probability a user will choose a particular item. After agents decide which of the recommended items they will choose, their data are fed into the student model whose matrices have an arbitrary low rank $k'$ (not necessarily equal to $k=4$). We chose $k'=5$ in our simulations, although results are similar if we chose $k'=2$, but the fit to data is worse. While some matrix factorization models train with a ridge regression loss term \cite{Bell2007,funk2006netflix}, we choose a slightly easier approach: stochastic gradient decent (SGD) with early stopping. We split the available data at random with 80\% for training and 20\% for validation and apply SGD until the 20\% validation set's brier score error is minimized (where the initial conditions are the matrix weights from the previous timepoint). While this differs from some recommendation algorithms, previous work has shown that matrix factorization is a variant of dense neural networks that commonly implement this method \cite{Lesieur2017}, and there is a close quantitative connection between ridge regression and early stopping \cite{Gunasekar2018}. Finally, this method allows us to stop training early, which makes simulations run faster. The recommendation algorithm uses this model to recommend $r=1$ item to each agent following a strategy outlined below.

\subsection{Recommendation Algorithm Strategies}

We propose a number of realistic strategies to recommend items, including a greedy strategy (recommending the content the student model predicts will most likely be chosen), an $\epsilon$-greedy strategy, in which random unchosen content is recommended with probability $\epsilon$, and a random strategy, in which random unchosen content is recommended with equal probability. We compare these against the best case scenario, the oracle strategy, in which we unrealistically set the student model equal to the teacher model. The $\epsilon$-greedy strategy is inspired by reinforcement learning systems \cite{Kaelbling1996,Arulkumaran2017,Afsar2021}, where recommendations are usually built up from interactions between individuals and the system. In contrast to previous work \cite{Afsar2021}, however, we incorporate reinforcement learning strategies into training a collaborative filtering model.

\subsection{Simulation Parameters}
These recommendations are then stochastically chosen by agents following the teacher model probabilities, and the simulation repeats until all items have been recommended. The student model is first trained on a sparse initial set of data. More specifically, 0.1\% of data was initially sampled uniformly a random with values $R^{\text{data}}_{ij}=$ 0 or 1 depending on a Bernoulli distribution with probability ${R}^{\text{teacher}}_{ij}$. Many datasets, such as the Netflix Prize dataset, have a greater proportion of user-item pairs (roughly 1\% of all possible pairs \cite{Bell2007}). However, the user-item pairs were themselves recommended with Netflix's in-house algorithm that was trained on an even smaller set of data, which we assume is 0.1\% of all possible pairs. 

We run this model for $n=4000$ agents and $m=200$ simulated items with ten realizations for each value of $\beta$ (or five realizations for the random $\boldsymbol{\beta}$ teacher model). A realization is where we retrain the student model from scratch, starting with a random 0.1\% of all pairs. We also generate a new teacher model, but keeping the same teacher model does not significantly affect results. 
The ratio of agents to items was chosen to approximately correspond to that of the Netflix prize data \cite{Bell2007}, roughly 20 agents per item. Largely because of the number of times we fit the student model over the course of each simulation realization ($m=200$ times), and because matrix factorization takes up to $O(n\times m)$, the simulation takes $O(n\times m^2)$. This time complexity means that modeling our comparatively modest set of agents and items would take several computing days for one simulation realization if it were not run in parallel. We are able to finish all the simulations in this paper in roughly 1-2 weeks on three 48-logical-core servers using Python (see link to code in the Introduction).

\begin{figure*}[tbh!]
    \centering
    \includegraphics[width=0.7\linewidth]{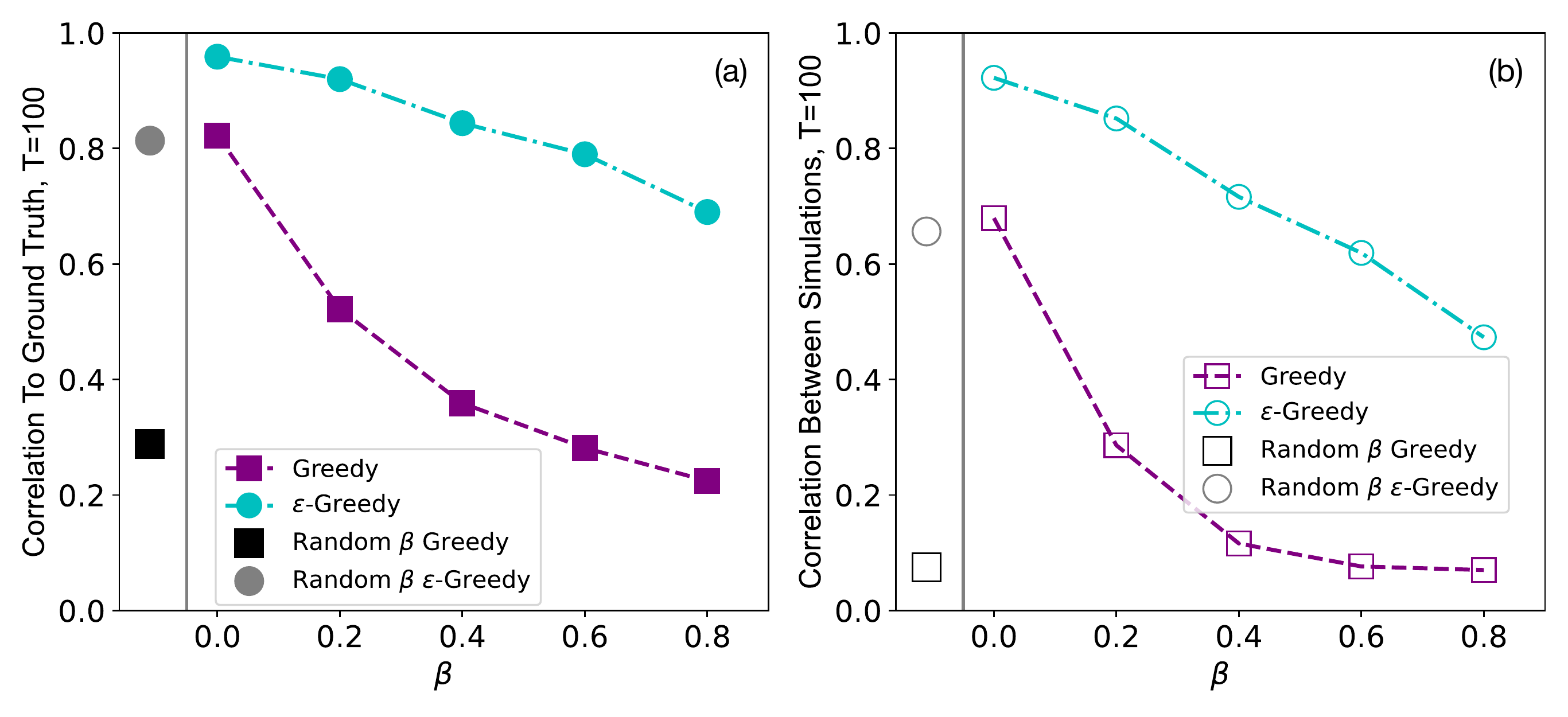}
    \caption{Student model instability. (a) The correlation between item popularity at timestep 100 and the teacher model probability an item would be chosen as a function of the human bias parameter, $\beta$. An alternative random $\boldsymbol{\beta}$ model, where the bias is uniformly distributed between 0 and 1 for each user-item pair, shows similar results. (b) The item popularity correlation between different realizations of the model after 100 timesteps. }
    \label{fig:Correl}
\end{figure*}

\section{Results}

We test whether the recommendation algorithm provides the agents with the items they want, whether items are ranked consistently, and finally how to improve recommendation algorithm stability.

We can gain intuition about how the greedy strategy affects the system in the limit that the student model's low-rank matrices are rank-one matrices ($k'=1$), which is even simpler than the simulation discussed in the rest of the paper ($k'=5$). If we want to recommend the top items to each agent, $i$, then we want to find item $j$ with the largest value in the student model, ${R}^{\text{student}}_{ij}= P'_i \times Q'_j$. However, in this case, the system recommends the same item to each agent because the relative ranking of items only depends on  $Q'_j$. This common ranking also implies that the recommendation system will only recommend popular content to agents rather than niche items agents may prefer. The homogeneity in recommendations and relationship between recommendations and popularity is seen in previous work on more realistic systems \cite{Chaney2018,Mansoury2020}, therefore even this simplified version of the simulation captures realism of more sophisticated systems. What is not captured in previous work, however, is that $\mathbf{Q'}$ would vary dramatically depending on the initial conditions, which implies item popularity instability: the same item could be very popular or unpopular in different simulation realizations by chance. The $\epsilon$-greedy strategy, in contrast to the greedy strategy, promotes random items, which we will show reduces the inequality of the system and helps the recommendation algorithm quickly find the most preferred content.

\subsection{Instability of Recommendation Systems}

Figure~\ref{fig:Correl} shows the stability and accuracy of the model. Figure~\ref{fig:Correl}a compares the popularity of items after 100 timesteps, when half of all recommendations are made, to the ground truth (popularity if all user-item pairs were fully sampled). We find that increasing the bias $\beta$ decreases the correlation between algorithm and ground truth popularity, therefore items that should be popular are not. Figure~\ref{fig:Correl}b, in contrast, shows that larger $\beta$ decreases the item popularity correlation between simulation realizations, implying greater item popularity instability. This is alike to previous work on crowdsource systems \cite{Burghardt2020,Salganik2006}, in which ranking items by a simple heuristic can drive some items to become popular by chance. Despite this finding, the $\epsilon$-greedy strategy creates much higher correlations between item popularity and the ground truth (Fig.~\ref{fig:Correl}a) and item popularities between each simulation realization (Fig.~\ref{fig:Correl}b). The new strategy, in other words, improves recommendation accuracy and reduces item popularity instability.

\begin{figure}[tb!]
    \centering
    \includegraphics[width=0.9\linewidth]{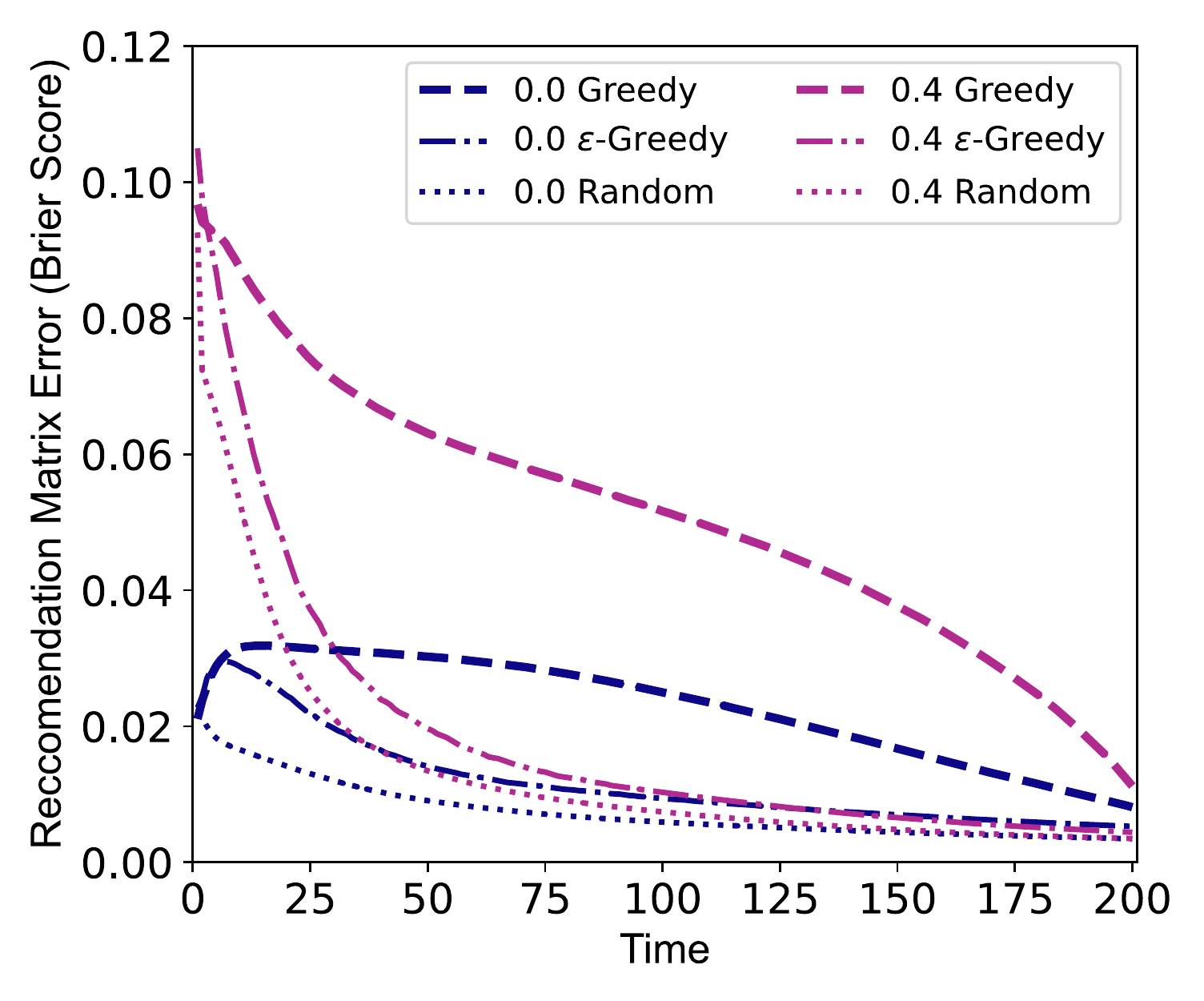}
    \caption{Model error for different algorithm strategies. We find that the Brier score (mean squared error between model prediction and agent decision) drops with time and is highest when the algorithm uses the greedy strategy for $\beta=0.0$, $0.4$. On the other hand, when the algorithm follows the  $\epsilon$-greedy strategy, the error drops drops dramatically, and is nearly as low as the random strategy (offering random items to agents).}
    \label{fig:Error}
\end{figure}
\begin{figure*}[tbh!]
    \centering
    \includegraphics[width=0.7\linewidth]{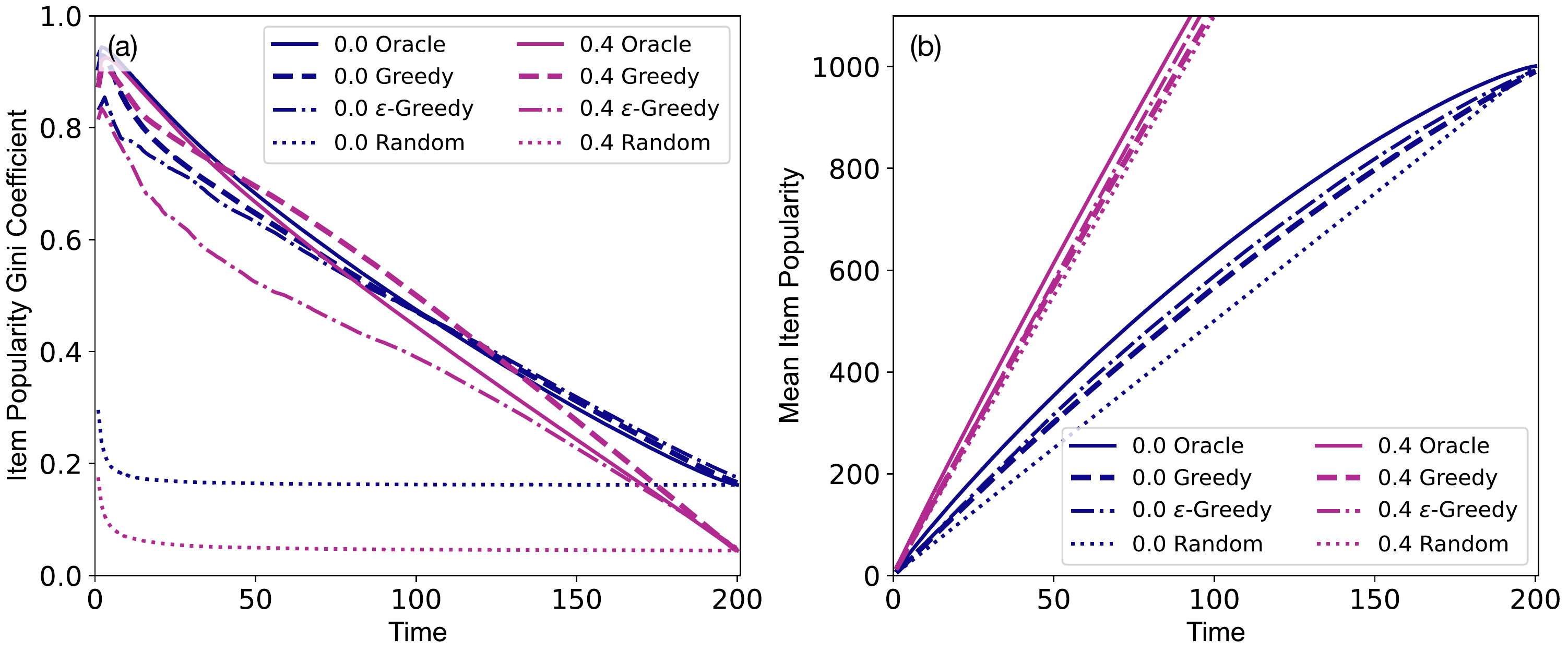}
    \caption{The evolution of item popularity. (a) Gini coefficient and (b) mean item popularity over time for $\beta=0.0$, $0.4$. Four different strategies are used for recommendation: oracle, greedy, $\epsilon$-greedy and random. Gini coefficient is generally lower for $\epsilon$-greedy and random strategies, and the $\epsilon$-greedy strategy makes more ideal recommendations, allowing mean item popularity to be higher than all but the oracle strategy.}
    \label{fig:Views}
\end{figure*}

The reason the greedy strategy performs poorly can be better understood when we plot student model error over time in Fig.~\ref{fig:Error}. We show that model error generally decreases with time, as expected, but the greedy strategy has error decreasing slowly with time. The $\epsilon$-greedy strategy enhances the student model through a representative sample of the user-item matrix. Error for this strategy therefore drops to a small fraction of the greedy strategy and is nearly as small as the error for the random strategy.

\subsection{Comparing Recommendation Quality}

Next, we compare the quality of recommendations by observing the items chosen over time in Fig.~\ref{fig:Views}. We show that the $\epsilon$-greedy strategy makes more diverse recommendations that are of higher quality on average than the random or greedy strategies. In Fig.~\ref{fig:Views}a, we show the Gini coefficient of item popularity, a proxy for item popularity inequality. If items were fully sampled, their Gini coefficient would be less than 0.2 ($T=200$ values on the right side of the figure). Under the idealized oracle strategy, the Gini coefficient is initially high (only a few of the best items are recommended) and steadily drops. The lower Gini coefficient for the $\epsilon$-greedy strategy is a product of more equal sampling. The greedy strategy Gini coefficient is, however, often higher than all alternative strategies, meaning some items are recommended far more frequently than they should be. 

\begin{figure*}[tbh!]
    \centering
    \includegraphics[width=0.95\linewidth]{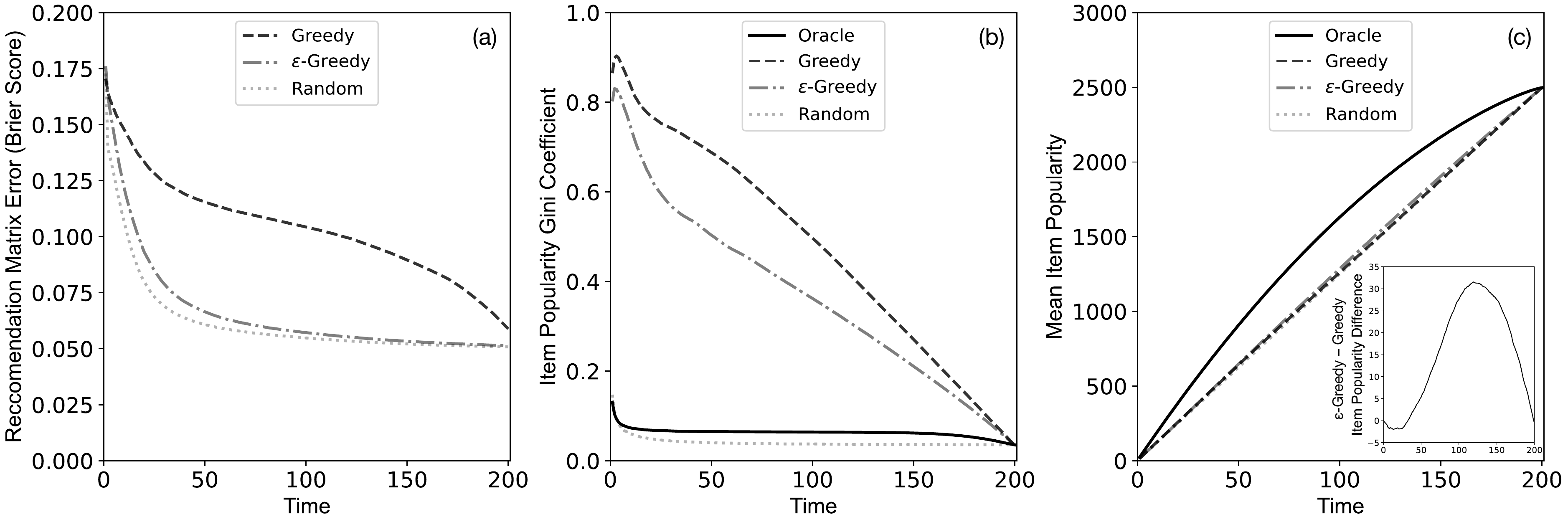}
    \caption{Robustness of results. (a) Brier score, (b) item popularity Gini coefficient, (c) mean item popularity over time for random $\boldsymbol{\beta}$ matrix. As in the previous experiments, the $\epsilon$-greedy strategy has lower Brier score and item popularity Gini coefficient than the greedy strategy, and slightly (but statistically significantly) higher mean item popularity. Difference in item popularity between $\epsilon$-greedy and greedy strategy over time is shown in the inset.}
    \label{fig:Robust}
\end{figure*}

Next, we plot the mean popularity over time in Fig.~\ref{fig:Views}b. If the mean popularity of items is high early on then agents are recommended items they really like. The mean popularity of items is highest in the oracle strategy because we know exactly what items users are most likely to choose and recommend those first. We find agents choose more items at any timepoint with the $\epsilon$-greedy strategy than alternatives implying it is the best non-oracle strategy. The random recommendations within the $\epsilon$-greedy strategy trains the student model better, and this in turn creates better recommendations on average. 

\subsection{Robustness of Results}

The simulations shown in Figs.~\ref{fig:Error} \&~\ref{fig:Views} make several simplifying assumptions, including a $\boldsymbol{\beta}$ matrix whose elements are all the same. Instead $\boldsymbol{\beta}$ elements could be very different, which we model as values that are independent and uniformly distributed between 0 and 1. We show in Fig.~\ref{fig:Correl} that this does not change our finding that the greedy strategy is less stable than the $\epsilon$-greedy strategy. We similarly show that the model error and item popularity is qualitatively similar in Fig.~\ref{fig:Robust}. Namely, in Fig.~\ref{fig:Robust}a, we show that the greedy method has consistently higher model error than the $\epsilon$-greedy strategy. Similarly, the Gini coefficient (Fig.~\ref{fig:Robust}b) is higher and the mean popularity over time is very slightly, but statistically significantly, lower in Fig.~\ref{fig:Robust}c (shown more clearly in the inset, which shows the difference in item popularity over time). To test the statistical significance, we take the mean popularity difference across all timesteps between the greedy and $\epsilon$-greedy strategies and compare the z-score of this difference (which is greater than $15$), making the $p$-value$<10^{-6}$. This result is approximately the same when we compare the mean difference for individual realizations or the mean item popularity across five realizations; results stated above are for mean popularity across five realizations. Unlike in earlier results, however, the teacher model is now poorly approximated as a product of two lower-rank matrices, therefore error is typically higher. The oracle strategy in turn has a lower Gini coefficient and higher mean popularity than alternative strategies.

\section{Discussion \& Conclusions}

In conclusion, we develop a teacher-student framework to understand the accuracy and stability of collaborative filtering models as they interact with agents. The simulations demonstrate that the greedy strategy produces unstable and inaccurate predictions over time. Namely, the recommendation algorithm recommends a small set of items (leading to a high popularity Gini coefficient) and this leads to higher error. In contrast, the $\epsilon$-greedy strategy follows a less intuitive training regime in which random items are recommended to agents. This leads to better sampling (lower Gini coefficient), lower error, and \emph{more} items picked at any given time because the items recommended are what agents prefer to choose. Finally, the $\epsilon$-greedy strategy might force users out of filter bubbles by exposing them to a more diverse set of items. This potential should be explored in future work. This paper adds to growing literature on the instability of ranking systems \cite{Salganik2006, Burghardt2020}, but also gives greater insight into personalized ranking, and emergent properties, both desired and unintended, of these systems. For example, the sensitivity of the recommendation algorithm to initial conditions is reminiscent of chaotic systems, but future work is needed to test the relationship between these findings and non-linear dynamics or chaos theory.

\subsection{Limitations}

There are a number of potential limitations with the current method. First, our work must rely on synthetic data because we cannot know whether a user would choose an item they were not recommended in empirical data. Furthermore, assumptions built into the present simulation may not reflect the true human dynamics. For example, agents are the same age and equally active in the system. In reality, some agents may be much older and have much more data than others. In addition, the order items are recommended could affect agent decisions. For example, users might not buy a vinyl record unless they are first recommended a record player. Moreover, we use a simple student model to recommend items, but newer and more sophisticated collaborative filtering methods could be offered. 

\subsection{Ethical Considerations}
The present simulations offer policy suggestions to improve recommendations by combining ideas from collaborative filtering with reinforcement learning. This current work does not, however, explore the potential adverse effects of recommendation systems, such as filter bubbles \cite{Bessi2016}. Items recommended by the $\epsilon$-greedy strategy could be of low-quality or promote harm, while we would expect such items are screened out when recommendation algorithms promote what agents should like.  That said, this method actively fights filter bubbles by offering items outside of the user's expected preferences, and a variation of this strategy, such as recommendations among a cleaned or popular set of items could provide users better and more diverse items.

\subsection{Future Work}
Additional realism, such as agents arriving or leaving the system, could easily be incorporated into the simulation. The present work is a baseline that researchers can modify for specific systems. Additional features should be explored, however, including polarization, in which people preferentially pick one type of content over another. While there has been a growing interest in algorithmic polarization \cite{Sirbu2019}, the dynamic interaction between agents and trained models should be explored in greater depth, especially if it can drive people away from echo chambers. Moreover, the teacher-student methodology in this paper can be extended to audit other socio-technical systems such as predictive policing \cite{Ensign2018}, bail \cite{kleinberg2016inherent}, or banking loans, whose data is known to be intrinsically biased \cite{DAmour2020}. The instability we could measure is, for example, who gets loans or goes to jail. If this varies due to the simulation realization and not the intrinsic features of the people, we could quantify, and find ways to address, the algorithm's instability. 

\section*{Acknowledgements}
Research was supported by  DARPA under award \# HR001121C0169. 

\section*{Conflicts of Interest}
The authors declare no conflicts of interest.


\begin{thebibliography}{38}
\providecommand{\natexlab}[1]{#1}
\providecommand{\url}[1]{\texttt{#1}}
\providecommand{\urlprefix}{URL }
\expandafter\ifx\csname urlstyle\endcsname\relax
  \providecommand{\doi}[1]{doi:\discretionary{}{}{}#1}\else
  \providecommand{\doi}{doi:\discretionary{}{}{}\begingroup
  \urlstyle{rm}\Url}\fi

\bibitem[{Afsar, Crump, and Far(2021)}]{Afsar2021}
Afsar, M.~M.; Crump, T.; and Far, B. 2021.
\newblock Reinforcement learning based recommender systems: A survey.
\newblock \emph{arXiv preprint arXiv:2101.06286} .

\bibitem[{Angwin et~al.(2016)Angwin, Larson, Mattu, and
  Kirchner}]{angwin2016machine}
Angwin, J.; Larson, J.; Mattu, S.; and Kirchner, L. 2016.
\newblock Machine bias. ProPublica.
\newblock \emph{See https://www. propublica.
  org/article/machine-bias-risk-assessments-in-criminal-sentencing} .

\bibitem[{Arulkumaran et~al.(2017)Arulkumaran, Deisenroth, Brundage, and
  Bharath}]{Arulkumaran2017}
Arulkumaran, K.; Deisenroth, M.~P.; Brundage, M.; and Bharath, A.~A. 2017.
\newblock A Brief Survey of Deep Reinforcement Learning.
\newblock \emph{arXiv preprint arXiv:1708.05866} .

\bibitem[{Balabanovi\'{c} and Shoham(1997)}]{Balabanovic1997}
Balabanovi\'{c}, M.; and Shoham, Y. 1997.
\newblock Fab: Content-Based, Collaborative Recommendation.
\newblock \emph{Commun. ACM} 40(3): 66–72.
\newblock ISSN 0001-0782.
\newblock \doi{10.1145/245108.245124}.
\newblock \urlprefix\url{https://doi.org/10.1145/245108.245124}.

\bibitem[{Bell and Koren(2007)}]{Bell2007}
Bell, R.~M.; and Koren, Y. 2007.
\newblock Lessons from the Netflix Prize Challenge.
\newblock \emph{SIGKDD Explor. Newsl.} 9(2): 75–79.
\newblock ISSN 1931-0145.
\newblock \doi{10.1145/1345448.1345465}.
\newblock \urlprefix\url{https://doi.org/10.1145/1345448.1345465}.

\bibitem[{Bessi et~al.(2016)Bessi, Zollo, Del~Vicario, Puliga, Scala,
  Caldarelli, Uzzi, and Quattrociocchi}]{Bessi2016}
Bessi, A.; Zollo, F.; Del~Vicario, M.; Puliga, M.; Scala, A.; Caldarelli, G.;
  Uzzi, B.; and Quattrociocchi, W. 2016.
\newblock Users Polarization on Facebook and Youtube.
\newblock \emph{PLOS ONE} 11(8): 1--24.
\newblock \doi{10.1371/journal.pone.0159641}.
\newblock \urlprefix\url{https://doi.org/10.1371/journal.pone.0159641}.

\bibitem[{Biancalana et~al.(2011)Biancalana, Gasparetti, Micarelli, Miola, and
  Sansonetti}]{biancalana2011context}
Biancalana, C.; Gasparetti, F.; Micarelli, A.; Miola, A.; and Sansonetti, G.
  2011.
\newblock Context-aware movie recommendation based on signal processing and
  machine learning.
\newblock In \emph{Proceedings of the 2nd Challenge on Context-Aware Movie
  Recommendation}, 5--10. ACM.

\bibitem[{Burghardt et~al.(2017)Burghardt, Alsina, Girvan, Rand, and
  Lerman}]{MyopiaCrowd}
Burghardt, K.; Alsina, E.~F.; Girvan, M.; Rand, W.; and Lerman, K. 2017.
\newblock The Myopia of Crowds: A Study of Collective Evaluation on Stack
  Exchange.
\newblock \emph{PLOS ONE} 12(3): e0173610.

\bibitem[{Burghardt et~al.(2020)Burghardt, Hogg, D'Souza, Lerman, and
  Posfai}]{Burghardt2020}
Burghardt, K.; Hogg, T.; D'Souza, R.; Lerman, K.; and Posfai, M. 2020.
\newblock Origins of Algorithmic Instabilities in Crowdsourced Ranking.
\newblock \emph{Proc. ACM Hum.-Comput. Interact.} 4(CSCW2).
\newblock \doi{10.1145/3415237}.
\newblock \urlprefix\url{https://doi.org/10.1145/3415237}.

\bibitem[{Chaney, Stewart, and Engelhardt(2018)}]{Chaney2018}
Chaney, A. J.~B.; Stewart, B.~M.; and Engelhardt, B.~E. 2018.
\newblock How Algorithmic Confounding in Recommendation Systems Increases
  Homogeneity and Decreases Utility.
\newblock In \emph{Proceedings of the 12th ACM Conference on Recommender
  Systems}, RecSys '18, 224–232. New York, NY, USA: Association for Computing
  Machinery.
\newblock ISBN 9781450359016.
\newblock \doi{10.1145/3240323.3240370}.
\newblock \urlprefix\url{https://doi.org/10.1145/3240323.3240370}.

\bibitem[{Conley and Perry(2013)}]{conley2013does}
Conley, K.; and Perry, D. 2013.
\newblock How does he saw me? a recommendation engine for picking heroes in
  dota 2.
\newblock \emph{Np, nd Web} 7.

\bibitem[{D'Amour et~al.(2020)D'Amour, Srinivasan, Atwood, Baljekar, Sculley,
  and Halpern}]{DAmour2020}
D'Amour, A.; Srinivasan, H.; Atwood, J.; Baljekar, P.; Sculley, D.; and
  Halpern, Y. 2020.
\newblock Fairness is Not Static: Deeper Understanding of Long Term Fairness
  via Simulation Studies.
\newblock In \emph{Proceedings of the 2020 Conference on Fairness,
  Accountability, and Transparency}, FAT* '20, 525–534. New York, NY, USA:
  Association for Computing Machinery.
\newblock ISBN 9781450369367.
\newblock \doi{10.1145/3351095.3372878}.
\newblock \urlprefix\url{https://doi.org/10.1145/3351095.3372878}.

\bibitem[{Ensign et~al.(2018)Ensign, Friedler, Neville, Scheidegger, and
  Venkatasubramanian}]{Ensign2018}
Ensign, D.; Friedler, S.~A.; Neville, S.; Scheidegger, C.; and
  Venkatasubramanian, S. 2018.
\newblock Runaway Feedback Loops in Predictive Policing.
\newblock In Friedler, S.~A.; and Wilson, C., eds., \emph{Proceedings of the
  1st Conference on Fairness, Accountability and Transparency}, volume~81 of
  \emph{Proceedings of Machine Learning Research}, 160--171. PMLR.
\newblock \urlprefix\url{https://proceedings.mlr.press/v81/ensign18a.html}.

\bibitem[{Funk(2006)}]{funk2006netflix}
Funk, S. 2006.
\newblock Netflix update: Try this at home.

\bibitem[{Ge et~al.(2020)Ge, Zhao, Zhou, Pei, Sun, Ou, and Zhang}]{Ge2020}
Ge, Y.; Zhao, S.; Zhou, H.; Pei, C.; Sun, F.; Ou, W.; and Zhang, Y. 2020.
\newblock Understanding Echo Chambers in E-Commerce Recommender Systems.
\newblock In \emph{Proceedings of the 43rd International ACM SIGIR Conference
  on Research and Development in Information Retrieval}, SIGIR '20,
  2261–2270. New York, NY, USA: Association for Computing Machinery.
\newblock ISBN 9781450380164.
\newblock \doi{10.1145/3397271.3401431}.
\newblock \urlprefix\url{https://doi.org/10.1145/3397271.3401431}.

\bibitem[{Glenski et~al.(2018)Glenski, Stoddard, Resnick, and
  Weninger}]{Glenski2018}
Glenski, M.; Stoddard, G.; Resnick, P.; and Weninger, T. 2018.
\newblock GuessTheKarma: A Game to Assess Social Rating Systems.
\newblock \emph{Proc. ACM Hum.-Comput. Interact.} 2(CSCW).
\newblock \doi{10.1145/3274328}.
\newblock \urlprefix\url{https://doi.org/10.1145/3274328}.

\bibitem[{Gunasekar et~al.(2018)Gunasekar, Woodworth, Bhojanapalli, Neyshabur,
  and Srebro}]{Gunasekar2018}
Gunasekar, S.; Woodworth, B.; Bhojanapalli, S.; Neyshabur, B.; and Srebro, N.
  2018.
\newblock Implicit Regularization in Matrix Factorization.
\newblock In \emph{2018 Information Theory and Applications Workshop (ITA)},
  1--10.
\newblock \doi{10.1109/ITA.2018.8503198}.

\bibitem[{He et~al.(2017)He, Liao, Zhang, Nie, Hu, and Chua}]{He2017}
He, X.; Liao, L.; Zhang, H.; Nie, L.; Hu, X.; and Chua, T.-S. 2017.
\newblock Neural Collaborative Filtering.
\newblock In \emph{Proceedings of the 26th International Conference on World
  Wide Web}, WWW '17, 173–182. Republic and Canton of Geneva, CHE:
  International World Wide Web Conferences Steering Committee.
\newblock ISBN 9781450349130.
\newblock \doi{10.1145/3038912.3052569}.
\newblock \urlprefix\url{https://doi.org/10.1145/3038912.3052569}.

\bibitem[{Jabbari et~al.(2017)Jabbari, Joseph, Kearns, Morgenstern, and
  Roth}]{Jabbari2017}
Jabbari, S.; Joseph, M.; Kearns, M.; Morgenstern, J.; and Roth, A. 2017.
\newblock Fairness in Reinforcement Learning.
\newblock In Precup, D.; and Teh, Y.~W., eds., \emph{Proceedings of the 34th
  International Conference on Machine Learning}, volume~70 of \emph{Proceedings
  of Machine Learning Research}, 1617--1626. PMLR.
\newblock \urlprefix\url{https://proceedings.mlr.press/v70/jabbari17a.html}.

\bibitem[{Joseph et~al.(2016)Joseph, Kearns, Morgenstern, and
  Roth}]{Joseph2016}
Joseph, M.; Kearns, M.; Morgenstern, J.~H.; and Roth, A. 2016.
\newblock Fairness in Learning: Classic and Contextual Bandits.
\newblock In Lee, D.; Sugiyama, M.; Luxburg, U.; Guyon, I.; and Garnett, R.,
  eds., \emph{Advances in Neural Information Processing Systems}, volume~29.
  Curran Associates, Inc.
\newblock
  \urlprefix\url{https://proceedings.neurips.cc/paper/2016/file/eb163727917cbba1eea208541a643e74-Paper.pdf}.

\bibitem[{Kaelbling, Littman, and Moore(1996)}]{Kaelbling1996}
Kaelbling, L.~P.; Littman, M.~L.; and Moore, A.~W. 1996.
\newblock Reinforcement Learning: A Survey.
\newblock \emph{JAIR} 4: 237--285.

\bibitem[{Keith~Burghardt and Lerman(2018)}]{Burghardt2018}
Keith~Burghardt, T.~H.; and Lerman, K. 2018.
\newblock Quantifying the Impact of Cognitive Biases in Crowdsourcing.
\newblock In \emph{Proceedings of The 12th International {AAAI} Conference on
  Web and Social Media ({ICWSM}-18)}. AAAI.

\bibitem[{Kim et~al.(2016)Kim, Park, Oh, Lee, and Yu}]{kim2016convolutional}
Kim, D.; Park, C.; Oh, J.; Lee, S.; and Yu, H. 2016.
\newblock Convolutional matrix factorization for document context-aware
  recommendation.
\newblock In \emph{Proceedings of the 10th ACM Conference on Recommender
  Systems}, 233--240. ACM.

\bibitem[{Kleinberg, Mullainathan, and Raghavan(2016)}]{kleinberg2016inherent}
Kleinberg, J.; Mullainathan, S.; and Raghavan, M. 2016.
\newblock Inherent trade-offs in the fair determination of risk scores.
\newblock \emph{arXiv preprint arXiv:1609.05807} .

\bibitem[{Koren, Bell, and Volinsky(2009)}]{Koren2009}
Koren, Y.; Bell, R.; and Volinsky, C. 2009.
\newblock Matrix Factorization Techniques for Recommender Systems.
\newblock \emph{Computer} 42(8): 30--37.
\newblock \doi{10.1109/MC.2009.263}.

\bibitem[{Lampinen and Ganguli(2018)}]{Lampinen2018}
Lampinen, A.~K.; and Ganguli, S. 2018.
\newblock An analytic theory of generalization dynamics and transfer learning
  in deep linear networks.
\newblock \emph{arXiv preprint arXiv:1809.10374} .

\bibitem[{Lerman and Hogg(2014)}]{lerman14as}
Lerman, K.; and Hogg, T. 2014.
\newblock Leveraging position bias to improve peer recommendation.
\newblock \emph{PLOS ONE} 9(6): e98914.

\bibitem[{Lesieur, Krzakala, and Zdeborov{\'{a}}(2017)}]{Lesieur2017}
Lesieur, T.; Krzakala, F.; and Zdeborov{\'{a}}, L. 2017.
\newblock Constrained low-rank matrix estimation: phase transitions,
  approximate message passing and applications.
\newblock \emph{Journal of Statistical Mechanics: Theory and Experiment}
  2017(7): 073403.
\newblock \doi{10.1088/1742-5468/aa7284}.
\newblock \urlprefix\url{https://doi.org/10.1088/1742-5468/aa7284}.

\bibitem[{Lu and Yang(2016)}]{Lu2016}
Lu, Z.; and Yang, Q. 2016.
\newblock Partially Observable Markov Decision Process for Recommender Systems.
\newblock \emph{arXiv preprint arXiv:1608.07793} .

\bibitem[{Mansoury et~al.(2020)Mansoury, Abdollahpouri, Pechenizkiy, Mobasher,
  and Burke}]{Mansoury2020}
Mansoury, M.; Abdollahpouri, H.; Pechenizkiy, M.; Mobasher, B.; and Burke, R.
  2020.
\newblock Feedback Loop and Bias Amplification in Recommender Systems.
\newblock In \emph{Proceedings of the 29th ACM International Conference on
  Information \& Knowledge Management}, CIKM '20, 2145–2148. New York, NY,
  USA: Association for Computing Machinery.
\newblock ISBN 9781450368599.
\newblock \doi{10.1145/3340531.3412152}.
\newblock \urlprefix\url{https://doi.org/10.1145/3340531.3412152}.

\bibitem[{Portugal, Alencar, and Cowan(2018)}]{Portugal2018}
Portugal, I.; Alencar, P.; and Cowan, D. 2018.
\newblock The use of machine learning algorithms in recommender systems: A
  systematic review.
\newblock \emph{Expert Systems with Applications} 97: 205--227.
\newblock ISSN 0957-4174.
\newblock \doi{https://doi.org/10.1016/j.eswa.2017.12.020}.
\newblock
  \urlprefix\url{https://www.sciencedirect.com/science/article/pii/S0957417417308333}.

\bibitem[{Salganik, Dodds, and Watts(2006)}]{Salganik2006}
Salganik, M.~J.; Dodds, P.~S.; and Watts, D.~J. 2006.
\newblock Experimental Study of Inequality and Unpredictability in an
  Artificial Cultural Market.
\newblock \emph{Science} 311(5762): 854--856.
\newblock \doi{10.1126/science.1121066}.

\bibitem[{Schafer et~al.(2007)Schafer, Frankowski, Herlocker, and
  Sen}]{Schafer2007}
Schafer, J.~B.; Frankowski, D.; Herlocker, J.; and Sen, S. 2007.
\newblock \emph{Collaborative Filtering Recommender Systems}, 291--324.
\newblock Berlin, Heidelberg: Springer Berlin Heidelberg.
\newblock ISBN 978-3-540-72079-9.
\newblock \doi{10.1007/978-3-540-72079-9_9}.
\newblock \urlprefix\url{https://doi.org/10.1007/978-3-540-72079-9_9}.

\bibitem[{Sinha, Gleich, and Ramani(2016)}]{Sinha2016}
Sinha, A.; Gleich, D.~F.; and Ramani, K. 2016.
\newblock Deconvolving Feedback Loops in Recommender Systems.
\newblock In Lee, D.; Sugiyama, M.; Luxburg, U.; Guyon, I.; and Garnett, R.,
  eds., \emph{Advances in Neural Information Processing Systems}, volume~29.
  Curran Associates, Inc.
\newblock
  \urlprefix\url{https://proceedings.neurips.cc/paper/2016/file/962e56a8a0b0420d87272a682bfd1e53-Paper.pdf}.

\bibitem[{Sirbu et~al.(2019)Sirbu, Pedreschi, Giannotti, and
  Kert{\'e}sz}]{Sirbu2019}
Sirbu, A.; Pedreschi, D.; Giannotti, F.; and Kert{\'e}sz, J. 2019.
\newblock Algorithmic bias amplifies opinion fragmentation and polarization: A
  bounded confidence model.
\newblock \emph{PLOS ONE} 14(3): 1--20.
\newblock \doi{10.1371/journal.pone.0213246}.
\newblock \urlprefix\url{https://doi.org/10.1371/journal.pone.0213246}.

\bibitem[{Taghipour, Kardan, and Ghidary(2007)}]{Taghipour2007}
Taghipour, N.; Kardan, A.; and Ghidary, S.~S. 2007.
\newblock Usage-Based Web Recommendations: A Reinforcement Learning Approach.
\newblock In \emph{Proceedings of the 2007 ACM Conference on Recommender
  Systems}, RecSys '07, 113–120. New York, NY, USA: Association for Computing
  Machinery.
\newblock ISBN 9781595937308.
\newblock \doi{10.1145/1297231.1297250}.
\newblock \urlprefix\url{https://doi.org/10.1145/1297231.1297250}.

\bibitem[{Tan, Guo, and Li(2008)}]{Tan2008}
Tan, H.; Guo, J.; and Li, Y. 2008.
\newblock E-learning Recommendation System.
\newblock In \emph{2008 International Conference on Computer Science and
  Software Engineering}, volume~5, 430--433.
\newblock \doi{10.1109/CSSE.2008.305}.

\bibitem[{Trnecka and Trneckova(2021)}]{Trnecka2021}
Trnecka, M.; and Trneckova, M. 2021.
\newblock Model order selection for approximate Boolean matrix factorization
  problem.
\newblock \emph{Knowledge-Based Systems} 227: 107184.
\newblock ISSN 0950-7051.
\newblock \doi{https://doi.org/10.1016/j.knosys.2021.107184}.
\newblock
  \urlprefix\url{https://www.sciencedirect.com/science/article/pii/S0950705121004469}.

\end{thebibliography}
\end{document}